\documentclass[multphys,vecphys]{svmult}

\usepackage{makeidx}         
\usepackage{graphicx}        
\usepackage{multicol}        
\usepackage[bottom]{footmisc}
\RequirePackage{natbib}

\makeindex             

\newcommand{\apj}{ApJ}

\newcommand{\apjl}{ApJL}
\newcommand{\mnras}{MNRAS}

\newcommand{\araa}{ARA\&A}

\begin{document}

\title*{AMR Simulations of the Cosmological Light Cone: SZE Surveys
of the Synthetic Universe}
\author{Eric~J.~Hallman\inst{1}, Brian~W.~O'Shea\inst{2}, Michael L. Norman\inst{3}, Rick Wagner\inst{3}, Jack~O.~Burns\inst{1}}
\authorrunning{Hallman, O'Shea, Norman, Wagner, Burns}
\institute{Center for Astrophysics and Space Astronomy, Department of Astrophysical \& Planetary Science, University of Colorado, Boulder, CO 80309 \and
Los Alamos National Laboratory, Los Alamos, NM 87501 \and
Center for Astrophysics and Space Sciences, University of
California-San Diego, 9500 Gilman Drive, La Jolla, CA 92093}
\titlerunning{SZE Surveys of the Synthetic Universe}
\maketitle

\section{Abstract}
\label{sec:1}
 We present preliminary results from simulated large sky coverage ($\sim$100 square degrees) Sunyaev-Zeldovich effect (SZE) cluster surveys
using the cosmological adaptive mesh refinement N-body/hydro code Enzo.
We have generated simulated light cones to match the resolution and
sensitivity of current and future SZE instruments. These
simulations are the most advanced calculations of their kind. The
simulated sky surveys allow a direct comparison of large N-body/hydro
cosmological simulations to current and pending sky surveys. Our synthetic surveys provide an indispensable guide for observers in the interpretation of
large area sky surveys, and will develop the tools necessary to
discriminate between models for cluster baryonic physics, and to
accurately determine cosmological parameters.
\section{Background and Method}
\label{sec:2}
Clusters of galaxies form from the highest peaks in the
primordial spectrum of density perturbations generated by inflation in
the early universe. They are the most massive virialized structures in
the universe, and as such are rare objects. 
The number density of galaxy clusters as a function of mass and
redshift is strongly dependent on a number of cosmological
parameters \cite{wang, haiman01}. Cluster survey yields
depend on the value of the minimum flux probed as a function of
redshift, the growth function of structure, and the redshift evolution
of the comoving volume element \cite{rosati}.  

The simulation used to generate the light cones described in this poster is
of a 512 Mpc/h comoving volume of the universe, with the following
cosmological parameters: ($\Omega_b$, $\Omega_{CDM}$, $\Omega_{\Lambda}$, h, n,
$\sigma_8$) = (0.04, 0.26, 0.7, 0.7, 1.0, 0.9).  The simulation was initialized
on a $512^3$ root grid
with $512^3$ dark matter particles, corresponding to a dark matter (baryon)
mass resolution of $7.2 \times 10^{10}$ ($1.1 \times 10^{10}$) $M_{\odot}$/h and an initial
comoving spatial resolution of 1 Mpc/h.  The simulation was then evolved to
z=0 using a maximum of 4 levels of adaptive mesh refinement.  This
simulation results in a higher dynamic range than achieved by any previous AMR
cosmological simulation representing such a large physical volume.

These light cone simulations are run with both dark
matter and baryons, unlike most previous similar studies. It has been shown in recent studies
using both simulations \cite[e.g.,][]{rasia06} and high
resolution X-ray 
observations of galaxy clusters \cite[e.g.,][]{utp_obs, 1E06}
that many clusters show strong 
departures from both equilibrium and isothermality. These deviations
can have a strong impact on both the observable and derived properties
of clusters. We have
  previously shown that deviations of factors of 10 or more are common
  in SZE and X-ray observables during even low mass ratio
  mergers. Thus, in order to properly simulate sky surveys, it
    is critically important to self-consistently include baryons in
    numerical simulations.

The light cone discussed here was generated from the
    above simulation using a stacking method similar to that used by
    \cite{dasilva}, but with 100 times the
    angular coverage of the best previous N-body+hydro light cone of \cite{springel}. We simulate
an SZE observation of a 100 square degree region of the sky, looking at the
corresponding comoving volume of the universe from z=0.1 to z=2.75.  We use
26 discrete volumes of $\delta_z \simeq 0.1$. We have generated SZE Compton $y$ parameter
images that are 2048 pixels on a side and cover 10 degrees x 10
degrees at a resolution of approximately 17.5 arcseconds/pixel.  The images
are then degraded using Gaussian smoothing to the resolution of several
experiments which will be providing SZE data in the very near future,
including APEX-SZ, the South Pole Telescope, the Planck Surveyor and the Atacama Cosmology Telescope.
\begin{figure}
\centering
\includegraphics[width=4.1in]{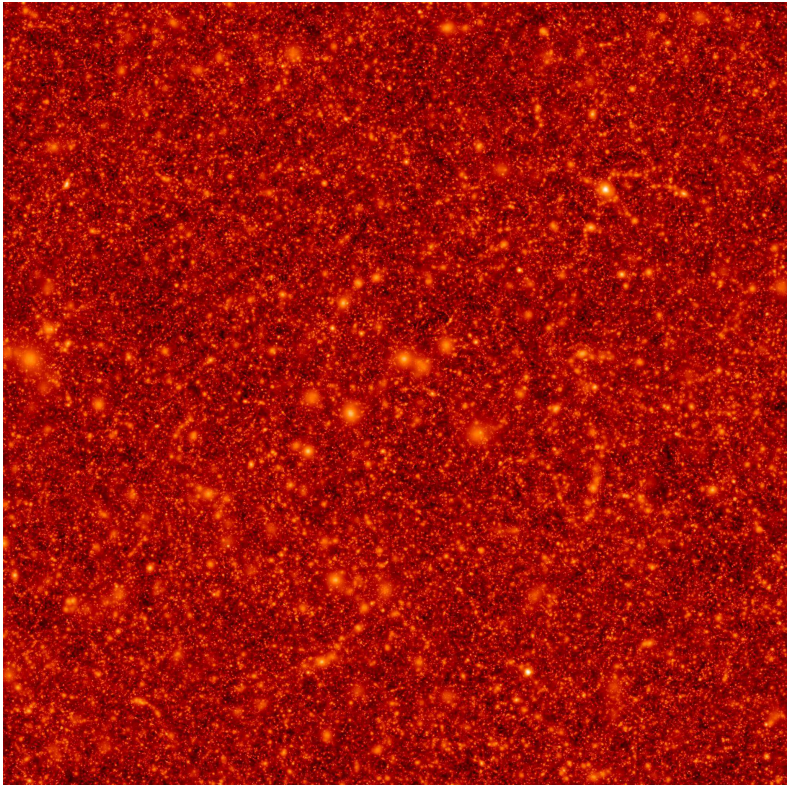}
\caption{10 $\times$ 10 degree
  synthetic SZE Compton $y$ survey image from Enzo AMR simulation of
  comoving 512 Mpc/h volume. The simulation contains 512$^3$ root grid
  zones and  512$^3$ dark matter particles with up to 4 levels of dynamic
  refinement. The image contains 2048$^2$ pixels for an
  angular resolution of $\sim$0.3 arcmin/pixel. Image contains
  structures from z=2.75 to z=0.1.}
\end{figure}

This work was supported in part by a grant from the U.S. National Science Foundation (AST-0407368).

\printindex
\end{document}